\documentclass[twocolumn,superscriptaddress,showpacs,preprintnumbers,
                                               amsmath,amssymb,prl]{revtex4}
\usepackage{graphicx}
\usepackage{dcolumn}
\usepackage{bibentry}

\begin{document}


\title{Revised value of the eighth-order electron $g\!-\!2$}

\author{T. Aoyama}
\affiliation{Theoretical Physics Laboratory,
Nishina Center, RIKEN, Wako, Saitama 351-0198, Japan }

\author{M. Hayakawa}
\affiliation{Department of Physics, Nagoya University, Nagoya, 
Aichi 464-8602, Japan}

\author{T. Kinoshita}
\affiliation{Newman Laboratory for Elementary-Particle Physics,    
Cornell University, Ithaca, New York 14853, U.S.A.}

\author{M. Nio}
\affiliation{Theoretical Physics Laboratory,
Nishina Center, RIKEN, Wako, Saitama 351-0198, Japan }

\date{\today}

\begin{abstract}
The contribution to the eighth-order anomalous magnetic 
moment ($g\! - \! 2$) of the electron from a set of diagrams 
without closed lepton loops
is recalculated using a new FORTRAN code
generated by an automatic code generator.
Comparing the contributions of individual diagrams  
of old and new calculations,
we found an inconsistency in the old treatment
of infrared subtraction terms in two diagrams.
Correcting this error leads to the 
revised value $-1.9144~(35) (\alpha/\pi)^4$ 
for the eighth-order term.
This theoretical change induces the shift of the inverse of the fine
structure constant by $ -6.41180(73)\times 10^{-7}$.

\end{abstract}

\pacs{13.40.Em,14.60.Cd,12.20.Ds,06.20.Jr}

\newcommand{\AllFiguresOfEighthOrderSetV}[1]{
\begin{figure}[htbp!]
\includegraphics*[width=3.25in]{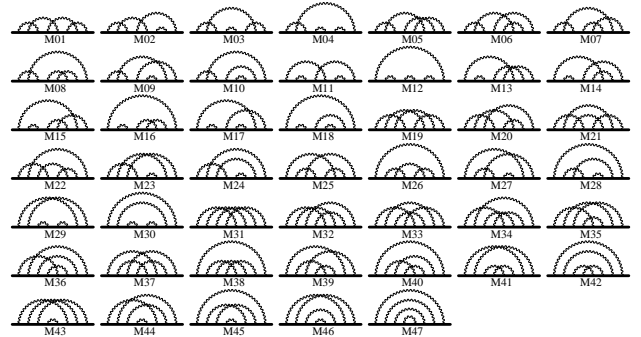}
\caption{Eighth-order Group V diagrams. 47 self-energy-like 
diagrams of $M_{01}$ -- $M_{47}$ 
represent 518 vertex diagrams.} 
\label{fig:EighthOrderSetV}
\end{figure}
}

\newcommand{\ComparFirstTable}{
\begin{table*}
\caption{
Comparison of the numerical calculation of $M_{01}$--$M_{24}$  of 
the eighth-order
Group V diagrams. The second column shows the analytic expression 
of the difference of old and new calculations
of the magnetic moment. The third column, value $A$, is obtained 
by plugging lower-order renormalization constants, such as 
$\Delta M_{4a}, \Delta L_{4s}$ 
into the expression in the second column. 
The fourth column, value $B$, lists 
the numerical values of $\Delta M^{\rm old} - \Delta M^{\rm new}$.
The fifth column is the difference $A-B$. If both numerical calculations 
are correct,
$A-B$ must vanish within the numerical uncertainty.
In evaluating $\Delta M^{\rm new}$ the double precision is used for the 
diagrams
without a self-energy subdiagram, while 
the quadruple precision is used for the reminder.
\label{Table:M01M24}
}
\newcolumntype{.}{D{.}{.}{8}}
\begin{ruledtabular}
\begin{tabular}{cl...}
\multicolumn{1}{c}{{\rm Diagram}} &
\multicolumn{1}{c}{{\rm difference}}&
\multicolumn{1}{c}{{\rm value}~$A$} &
\multicolumn{1}{c}{{\rm value}~$B$} &
\multicolumn{1}{c}{{$A-B$}} \\
\hline
$M_{01}$  &  0              
        &  0              &  -0.0129( 47)  &    0.0129( 47)  \\
$M_{02}$  & $2\Delta L_{6f1} M_2$                 
        &   -0.0066(  3)  &   0.0018(127)  &   -0.0084(127)  \\
$M_{03}$  & $\Delta L_{6f3} M_2$  
        &   -0.1132(  2)  &  -0.1055(100)  &   -0.0076(100)\\
$M_{04}$  & $2(\Delta L_{6d1} + \Delta L_{6d3} )M_2 $                 
        &    0.3338(  6)  &   0.3515(221)  &   -0.0177(221)  \\
$M_{05}$  &  0              
        &  0              &   0.0020( 28)  &   -0.0020( 28)  \\
$M_{06}$  &  0                      
        &  0              &  -0.0223( 61)  &    0.0223( 61)  \\
$M_{07}$  &  0                      
        &  0              &  -0.0102( 40)  &    0.0102( 40)  \\
$M_{08}$  & $2 (\Delta \delta m_{4a}\Delta M_{4a(1^*)}+\Delta L_{4c} 
   \Delta M_{4a}) $            
        &   -2.1809(  7)  &  -2.1773(163)  &   -0.0011(163)  \\
$M_{09}$  & $2\Delta L_{6f2} M_2 $                
        &    0.0805(  2)  &   0.0912(122)  &   -0.0106(122)  \\
$M_{10}$  & $2(\Delta \delta m_{4b}\Delta M_{4a(1^*)}+\Delta L_{6d2} M_2+  
                \Delta L_{4c}\Delta M_{4b}) $                   
        &   15.8899( 49)  &  15.8615(210)  &    0.0284(216)  \\
$M_{11}$  & $2\Delta L_{6d5} M_2 $              
        &    0.6948(  3)  &   0.6827(112)  &    0.0121(112)  \\
$M_{12}$  & $2(\Delta L_{6a1} + \Delta L_{6a3}) M_2 $                 
        &    1.2841(  0)  &   1.2875( 74)  &   -0.0034( 74)  \\
$M_{13}$  & $2\Delta L_{6h1} M_2 $                
        &   -0.4202(  4)  &  -0.4238( 48)  &    0.0036( 48)  \\
$M_{14}$  & $2\Delta L_{6g5} M_2 $                
        &    0.0892(  3)  &   0.0960( 95)  &   -0.0068( 95)  \\
$M_{15}$  & $2\Delta L_{6g1} M_2 $                
        &    0.0889(  3)  &   0.0893( 71)  &   -0.0004( 71)\\
$M_{16}$  & $2(\Delta \delta m_{4a}\Delta M_{4b(1^*)} 
       +\Delta L_{6c1}M_2 + \Delta L_{4s}\Delta M_{4a})$                  
        &   -2.6042(  6)  &  -2.6316(235)  &    0.0274(235)  \\
$M_{17}$  & $2(\Delta L_{6e1}  + \Delta L_{6d4}) M_2$                  
        &   -2.1183(  5)  &  -2.1010(189)  &   -0.0173(189)  \\
$M_{18}$  & $2 \{ \Delta \delta m_{4b}\Delta M_{4b(1^*)} 
      +\Delta L_{4s}\Delta M_{4b}+(\Delta L_{6b1} +\Delta L_{6a2}) M_2 \}$
        &   16.9690( 39)  &  17.1897(206)  &   -0.2207(210)  \\
$M_{19}$  &  0                      
        &    0            &   0.0002(  3)  &   -0.0002(  3)  \\
$M_{20}$  &  0                      
        &    0            &   0.0010( 17)  &   -0.0010( 17)  \\
$M_{21}$  &  0                      
        &    0            &   0.0003(  3)  &   -0.0003(  3)  \\
$M_{22}$  &  0                      
        &    0            &  -0.0090( 25)  &    0.0090( 25)  \\
$M_{23}$  & $2\Delta L_{6h2} M_2$                 
        &    0.0495(  3)  &   0.0438( 59)  &    0.0057( 59)  \\
$M_{24}$  & $2\Delta L_{6g2} M_2$                 
        &    0.0786(  2)  &   0.0945( 61)  &   -0.0158( 61)  \\ 
\end{tabular}
\end{ruledtabular}
\end{table*}
}
 
\newcommand{\ComparSecondTable}{
\begin{table*}
\caption{
Comparison of the numerical calculations of $M_{25}$-- $M_{47}$  of the 
eighth-order Group V diagrams. 
\label{Table:M25M47}
}
\newcolumntype{.}{D{.}{.}{8}}
\begin{ruledtabular}
\begin{tabular}{cl...}
\multicolumn{1}{c}{{\rm Diagram}} &
\multicolumn{1}{c}{{\rm difference} }&
\multicolumn{1}{c}{{\rm value}~$A$} &
\multicolumn{1}{c}{{\rm value}~$B$} &
\multicolumn{1}{c}{$A-B$} \\
\hline 
$M_{25}$  &  0                      
        &    0            &  -0.0031( 20)  &    0.0031( 20)  \\
$M_{26}$  & $\Delta \delta m_{6f}( M_{2^*} - M_{2^*}[I] ) $             
        &    2.5114(  4)  &   2.5369( 95)  &   -0.0255( 95)  \\
$M_{27}$  & $2\Delta L_{6g4} M_2                 $
        &   -0.0629(  2)  &  -0.0459( 90)  &   -0.0170( 90)  \\
$M_{28}$  & $2\{\Delta \delta m_{6d}( M_{2^*} - M_{2^*}[I] ) 
                          +\Delta L_{6c2} M_2 \} $              
        &   -7.5329(  6)  &  -7.5310(189)  &   -0.0020(189)  \\
$M_{29}$  & $2\Delta L_{6e2} M_2                $
        &   -0.2856(  3)  &  -0.2809(109)  &   -0.0047(109)  \\
$M_{30}$  & $\Delta \delta m_{6a}( M_{2^*} - M_{2^*}[I] ) 
                                 + 2\Delta L_{6b2} M_2   $          
        &    0.2768(  7)  &   0.2490(188)  &    0.0278(188) \\
$M_{31}$  &  0                      
        &    0            &   0.0007(  5)  &   -0.0007(  5)  \\
$M_{32}$  &  0                      
        &    0            &  -0.0024( 10)  &    0.0024( 10)  \\
$M_{33}$  &  0                      
        &    0            &   0.0001(  3)  &   -0.0001(  3)  \\
$M_{34}$  &  0                      
        &    0            &  -0.0010( 13)  &    0.0010( 13)  \\
$M_{35}$  &  0                      
        &    0            &   0.0001( 13)  &   -0.0001( 13)  \\
$M_{36}$  &  0                      
        &    0            &  -0.0027( 22)  &    0.0027( 22)  \\
$M_{37}$  &  0                      
        &    0            &   0.0004(  5)  &   -0.0004(  5)  \\
$M_{38}$  & $\Delta \delta m_{6h} ( M_{2^*} - M_{2^*}[I]$ )                       
        &   -0.9088(  3)  &  -0.9112( 40)  &    0.0025( 40)  \\
$M_{39}$  &  0                      
        &    0            &  -0.0031( 18)  &    0.0031( 18)  \\
$M_{40}$  & $2\Delta \delta m_{6g} ( M_{2^*} - M_{2^*}[I]$ )                       
        &    3.8271(  6)  &   3.8326( 71)  &   -0.0055( 71)  \\
$M_{41}$  & $\Delta \delta m_{4a} ( \Delta M_{4a(2^*)} )
                          +    \Delta L_{4x}\Delta M_{4a}     $           
        &    0.9809(  3)  &   0.9713( 83)  &    0.0096( 83)  \\
$M_{42}$  &  $ \Delta \delta m_{6c}( M_{2^*} - M_{2^*}[I] ) 
                                + \Delta L_{4l}\Delta M_{4a} $  
        &   -7.0216(  5)  &  -7.0202(114)  &   -0.0014(114)  \\
     &       $+ \Delta \delta m_{4a}      
                      \{\Delta M_{4b(2^*)}
             -\Delta \delta m_{2^*}(M_{2^*}-M_{2^*}[I])\}  $ 
                                                        \\       
$M_{43}$  & $\Delta L_{6h3} M_2 $                 
        &    0.4719(  2)  &   0.4703( 42)  &    0.0016( 42)  \\
$M_{44}$  & $2\Delta L_{6g3} M_2$                 
        &   -0.0751(  2)  &  -0.0499( 69)  &   -0.0253( 69)  \\
$M_{45}$  & $\Delta \delta m_{6e} ( M_{2^*} - M_{2^*}[I] )  
+ \Delta L_{6c3} M_2 $ 
        &   -0.0515(  6)  &  -0.0498( 90)  &   -0.0017( 90)  \\
$M_{46}$  &
         $\Delta\delta m_{4b}\Delta M_{4a(2^*)}+\Delta L_{6e3}M_2 
+ \Delta L_{4x}\Delta M_{4b}  $                 
        &   -7.9336( 22)  &  -7.9232( 86)  &   -0.0104(89)  \\
$M_{47}$ & $\Delta \delta m_{6b} ( M_{2^*} - M_{2^*}[I] )+ \Delta L_{6b3} M_2 + 
       \Delta L_{4l}\Delta M_{4b} $
        &   10.5868( 15)  &  10.5864(102)  &    0.0004(103)  \\ 
      & $+ \Delta\delta m_{4b} \{ \Delta M_{4b(2^*)}-\Delta\delta m_{2^*}
          (M_{2^*} - M_{2^*}[I] ) \} $ & & &  
\end{tabular}
\end{ruledtabular}
\end{table*}
}

\newcommand{\RconstSixthTable}{
\begin{table}
\caption{Finite renormalization constants used in 
Tables \ref{Table:M01M24} and  \ref{Table:M25M47}.
The validity of the sixth-order renormalization constants are checked
by comparing the sum $X_{LBD}\equiv \sum_{i=1}^5 \Delta L_{6xi} 
+ {1 \over 2} \Delta B_{6x} + 2 \Delta \delta m_{6x},~~
x=a,\cdots h$
to the previous $X_{LBD}$ values listed in Ref.~\cite{Kinoshita:2005zr}.
\label{Table:L6}
}
\newcolumntype{.}{D{.}{.}{10}}
\begin{ruledtabular}
\begin{tabular}{l.@{\hskip 2em}l.}
      $\Delta L_{6a1}$&    0.539589(67)&$\Delta L_{6b1}$&   -1.479347(244)  \\
      $\Delta L_{6a2}$&   -0.167211(81)&$\Delta L_{6b2}$&    0.582944(106)   \\
      $\Delta L_{6a3}$&    1.489038(142)&$\Delta L_{6b3}$&   -0.016344(73)  \\
      $\Delta L_{6c1}$&   -0.219311(148)&$\Delta L_{6e1}$&   -0.740890(373)  \\
      $\Delta L_{6c2}$&    0.071614(135)&$\Delta L_{6e2}$&   -0.285566(252)  \\
      $\Delta L_{6c3}$&   -0.551410(236)&$\Delta L_{6e3}$&   -0.141327(380)  \\
      $\Delta L_{6d1}$&    0.833454(402)&$\Delta L_{6g1}$&    0.088899(251)\\
      $\Delta L_{6d2}$&   -0.090653(141)&$\Delta L_{6g2}$&    0.078625(184)  \\
      $\Delta L_{6d3}$&   -0.499683(407)&$\Delta L_{6g3}$&   -0.075127(176)  \\
      $\Delta L_{6d4}$&   -1.377450 (287)&$\Delta L_{6g4}$&  -0.062906(155)  \\
      $\Delta L_{6d5}$&    0.694835 (227)&$\Delta L_{6g5}$&   0.089234(288)\\
      $\Delta L_{6f1}$&   -0.006638(212)& $\Delta L_{6h1}$&  -0.420233(330)\\
      $\Delta L_{6f2}$&    0.080534(139)& $\Delta L_{6h2}$&   0.049517(284)\\
      $\Delta L_{6f3}$&   -0.226304(227) &$\Delta L_{6h3}$&   0.943785(328)\\
       $\Delta \delta m_{6a}$  & -0.15309(34) &
      $\Delta \delta m_{6b}$  &  1.83775(25)  \\
      $\Delta \delta m_{6c}$  & -3.05039(22) &
      $\Delta \delta m_{6d}$  & -1.90114(15)  \\
      $\Delta \delta m_{6e}$  &  0.11210(25) &
      $\Delta \delta m_{6f}$  &  1.25572(19)  \\
      $\Delta \delta m_{6g}$  &  0.95677(13) &
      $\Delta \delta m_{6h}$  & -0.45439(11)    \\
      $\Delta L_{4c}$ &     0.003387(16) &
      $\Delta L_{4x}$ &    -0.481834(54)  \\
      $\Delta L_{4s}$ &     0.407633(20) &
      $\Delta L_{4l}$ &     0.124796(67)  \\
      $\Delta \delta m_{4a}$ &  -0.301485(61) &
      $\Delta \delta m_{4b}$ &   2.20777 (44) \\ 
      $\Delta M_{4a}$ &     0.218359(39)      & 
      $\Delta M_{4b}$ &    -0.187526(39)  \\
      $\Delta M_{4a(1^*)}$ &       3.61946(83)  &
      $\Delta M_{4a(2^*)}$ &      -3.60244(67) \\  
      $\Delta M_{4b(1^*)}$ &       4.25054(23)  &
      $\Delta M_{4b(2^*)}$ &       1.64475(10) \\
$\Delta M_2$         &        0.5  &
$\Delta M_{2^*}$     &        1    \\
$\Delta M_{2^*}[I]$  &       -1    &
$\Delta \delta m_{2^*}$ &     -0.75    \\ 
\end{tabular}
\end{ruledtabular}
\end{table}
}

\maketitle


\ComparFirstTable

The anomalous magnetic moment $(g-2)$ of the electron
has played a central role in testing the validity of QED  
\cite{Kusch:1947,Schwinger:1948iu}. 
Recently, a Harvard group measured the electron $g-2$ value \cite{Odom:2006gg} 
using a Penning Trap with a cylindrical cavity \cite{Brown:1985sa}.  
Their result for  $a_e \equiv (g\! - \! 2)/2$   
\cite{Odom:2006gg} 
\begin{equation}
a_e= 1~159~652~180.85(0.76)\times 10^{-12}~~~[0.66{\rm ppb}],
\label{aeHV06value}
\end{equation} 
has a 5.5 times smaller uncertainty than the best previous 
measurement \cite{VanDyck:1987ay}.

To match the precision of this measurement 
the theory of $g\!-\!2$ must 
include up to the eighth-order contribution of the QED perturbation theory  
\cite{Schwinger:1948iu,Petermann:1957, Sommerfield:1957, Kinoshita:1995,
Laporta:1996mq, Kinoshita:2005zr} 
as well as
the hadronic \cite{Jegerlehner:1996,Krause:1996}  and 
weak contributions \cite{Czarnecki:1996ww}. The tenth-order contribution
of QED $A_1^{(10)}(\alpha/\pi)^5$  might be relevant, 
but at present  it is not known.
As a temporary measure we adopt
the bound $A_1^{(10)} =  0~(3.8)$ proposed in
Ref.~\cite{CODATA:2002} to indicate a likely range
of value taken by $A_1^{(10)}$.
This will soon be replaced by an actual number
which is being evaluated right now
\cite{Aoyama:2005kf, Aoyama:2007IR, Kinoshita:2005sm}. 
Until then, the tenth-order term is the source of the largest theoretical
uncertainty of the electron $g\!-\!2$.
The next largest uncertainty comes from the numerical integration of the
eighth-order coefficient $A_1^{(8)}$
\cite{Kinoshita:2005zr}.

The purpose of this letter is to report the new value 
\begin{equation}
A_1^{(8)} = -1.914~4~(35)     
\label{newa8}
\end{equation}
obtained by combining the information derived from the previous
result \cite{Kinoshita:2005zr} and a new and independent 
evaluation of $A_1^{(8)}$
by means of FORTRAN codes generated by an automatic code 
generator ``gencodeN" \cite{Aoyama:2005kf,Aoyama:2007IR}.

$A_1^{(8)}$ receives contributions from 891 Feynman diagrams.
373 of them that have closed lepton loops had been 
calculated by more than
two independent methods 
\cite{Kinoshita:2005zr}.
The remaining 518 diagrams that have 
no closed lepton loop (called {\it q-type}) form one 
gauge invariant set (Group V).
In our formulation these diagrams are represented by
self-energy-like diagrams
related by the Ward-Takahashi identity. 
Taking the time reversal symmetry of QED into account,
518 vertex diagrams are amalgamated into 47 self-energy-like
diagrams shown in Fig.~\ref{fig:EighthOrderSetV}. 
Their integrands were carefully analyzed and checked by various means.
However, no independent check of calculation 
has been attempted until now.

\AllFiguresOfEighthOrderSetV

Technical progress in handling UV- and IR-divergences
has enabled us to generate the $N$th-order FORTRAN code
easily and swiftly \cite{Aoyama:2005kf,Aoyama:2007IR}.
Although ``gencodeN" was developed primarily
to handle the tenth-order term, we have applied it to 
fourth-, sixth- and eighth-order {\it q-type} diagrams
as part of the debugging effort. 
With the help of ``gencodeN"
eighth-order codes are generated easily.
However, their numerical evaluation by VEGAS \cite{vegas}
is quite nontrivial and requires huge computational resource. 
Numerical work has thus far reached
relative uncertainty of about
3 \% . Although this is more than an order of magnitude less accurate
than the uncertainty of the old calculation
\cite{Kinoshita:2005zr}, 
it is good enough for checking algebra of the old calculation.

Ultraviolet (UV) divergences of vertex and 
self-energy subdiagrams 
are removed by the $K$-operation 
\cite{Cvitanovic:1974uf, Cvitanovic:1974sv, Kinoshita:1990, Aoyama:2005kf},
which is identical with the old approach.
For diagrams containing self-energy subdiagrams, however, 
``gencodeN" treats UV-finite parts of self-energy subdiagrams
and IR divergences differently from the old approach 
\cite{Aoyama:2007IR}.

Comparison of the new ({\it still tentative}) and old calculations
has revealed an inconsistency in the treatment of the infrared
(IR) divergence in the latter, which is corrected in this letter.
Thus  we now have two independent evaluations of $A_1^{(8)}$.
Of course, 
much more numerical work is required to reach the precision
comparable to that of the old calculation.
Fortunately, correction terms themselves can be evaluated easily 
and very precisely as are shown in (\ref{M16_add}) and (\ref{M18_add}).

Finite integrals $\Delta M_i^{\rm old},~i=01,\cdots,47$, from 
the previous calculation are given in Ref.~\cite{Kinoshita:2005zr}.  
$\Delta M_i^{\rm new}$ are calculated 
using the programs generated by 
``gencodeN"\cite{Aoyama:2005kf,Aoyama:2007IR}. 
The numerical values corresponding
to $\Delta M_i^{\rm old} - \Delta M_i^{\rm new}$
are shown as value $B$ in Tables \ref{Table:M01M24} and \ref{Table:M25M47}.  
Since the diagrams without self-energy subdiagrams 
do not have IR divergence, 
$\Delta M_i^{\rm old}$ and
$\Delta M_i^{\rm new}$ should be identical. 
This is confirmed within the numerical precision of $\Delta M_i^{\rm new}$.
On the other hand, diagrams containing self-energy subdiagrams
have IR divergence. The new treatment of their contributions 
produces results different from those of Ref.~\cite{Kinoshita:2005zr}.
The difference
$\Delta M_i^{\rm old}- \Delta M_i^{\rm new}$ 
is listed symbolically
in the second column of Tables \ref{Table:M01M24} 
and \ref{Table:M25M47}.
Their numerical values are  calculated using the lower-order renormalization 
constants in Table \ref{Table:L6}  
and are shown as value $A$ in Tables \ref{Table:M01M24} and  \ref{Table:M25M47}.  
The difference of value $A$ and value $B$ is listed in 
the fifth columns of 
Tables \ref{Table:M01M24} and \ref{Table:M25M47}.
If both calculations are free from error, value $A$ and value $B$ 
must agree with
each other.

Tables \ref{Table:M01M24} and \ref{Table:M25M47} show that
``old" and ``new" calculations are in good agreement
for most diagrams. 
However, a large discrepancy $-0.221~(21)$ is found for the diagram $M_{18}$.  
Though no detectable discrepancy is found for $M_{16}$, 
it has a structure similar to $M_{18}$
and is somewhat simpler to analyze. Thus we examine here $M_{16}$ 
instead of $M_{18}$.

\ComparSecondTable

After an intense scrutiny of the programs of  $\Delta M_{16}^{\rm old}$ 
and $\Delta M_{16}^{\rm new}$, 
our attention was focused on
one of the IR subtraction terms of 
the finite term $\Delta M_{16}^{\rm old}$ 
\cite{Kinoshita:1981wm,Kinoshita:1990}:
\begin{eqnarray}
\Delta M_{16}^{\rm old} &\equiv& M_{16} - \sum_{f} \prod_{s\in f} 
{\mathbb K}_s M_{16} 
\nonumber \\
&-& I_{6c1} M_2 - \frac{1}{2} J_{6c} M_2 - I_{4s} \Delta M_{4a}
\nonumber \\
& -& \Delta \delta m_{4a} I_{4b(1^*)}
+I_{2^*} \Delta \delta m_{4a} M_2,
\end{eqnarray}
where $M_{16}$ is the bare amplitude, $\sum_{f} \prod_{s\in f} 
{\mathbb K}_s M_{14}$ are the UV counter terms
defined by the $K$-operations \cite{Cvitanovic:1974uf, Cvitanovic:1974sv, Kinoshita:1990, Aoyama:2005kf}, 
and the remainder are the IR subtraction terms.
By a term-by-term comparison, we found finally that
the IR subtraction term $I_{4b(1^*)}$ was the culprit.

Separation of an IR divergent part and a finite part
of an integral is arbitrary.  However, we must
keep track of what is treated as the IR divergent part.
In particular the IR subtraction term 
in $\Delta M_i$ and one used to calculate 
the residual renormalization must be identical.
All IR subtraction terms are summed up in the end,
which gives a finite
contribution as a part of the residual renormalization 
\cite{Kinoshita:1981wm, Kinoshita:1990, Aoyama:2007M8}.
What we found is that old FORTRAN codes 
of $I_{4b(1^*)}$ have different forms in
$\Delta M_{16}$  and in $\Delta M_{4b(1^*)}$.

If we use $I_{4b(1^*)}$  defined in 
Ref.~\cite{Kinoshita:1990} as a part of
$\Delta M_{4b(1^*)}$,
we must add the correction term
\begin{eqnarray}
\Delta M_{16}^{\rm add}  
         & \equiv &-2 \times {9 \over 4} \int (dz)_G { \delta m_{4a} [f_0] 
                   \over U^2 V^4 } \nonumber \\ 
        &&~~~~~~ \times z_2 A_2 (1-A_1)^3 (1-A_2) \nonumber \\
         & = & 0.029~437~8~(98) 
\label{M16_add}
\end{eqnarray}
to $\Delta M_{16}^{\rm old}$. 
The functions $A_i, U, V$ in Eq.~(\ref{M16_add}) are defined in 
the ${\mathbb I}_{1237}$ limit of the diagram $M_{16}$.
For precise definitions of these functions see 
Refs.~\cite{Cvitanovic:1974uf, Cvitanovic:1974sv,Kinoshita:1990, 
Aoyama:2005kf, Aoyama:2007M8}.
The overall factor 2 comes from the time-reversed diagram.
The value  (\ref{M16_add}) is smaller than the uncertainty 
of value $B$ for $M_{16}$.
Thus it is undetectable 
by direct comparison of values $A$ and $B$
until precision of $\Delta M_{16}^{\rm new}$ is improved. 

Analyzing the difference of $M_{18}^{\rm old}$ 
and $M_{18}^{\rm new}$ in the same manner, we found that 
the correction term is not small for $M_{18}$:
\begin{eqnarray}
\Delta M_{18}^{\rm add} &\equiv &-2\times {9 \over 4} \int (dz)_G
                  (1- {\mathbb K}_5)  \nonumber \\
                 &\times & \left \{ {  \delta m_{4b}[f_0] \over U^2 V^4}  
                   z_2 A_2 (1-A_1)^3 (1-A_2) \right \} 
                        \nonumber \\
                        & = & -0.215~542~(19),
\label{M18_add}
\end{eqnarray}
where all $A_i, U, V$ are defined in the ${\mathbb I}_{1237}$ 
limit of $M_{18}$.  Their
explicit forms are different from those of $M_{16}$.
The function $\delta m_{4a(b)}[f_0]$ in $M_{16(18)}^{\rm add}$ 
is related to the UV-finite part 
$\Delta \delta m_{4a(b)}$ of the
mass-renormalization constant.
If we add  $\Delta M_{18}^{\rm add}$ to $\Delta M_{18}^{\rm old}$, 
value $B$ of $M_{18}$ 
becomes $16.974~(21)$ and the difference between values $A$ and $B$
is reduced to $-0.005~(21)$, which is consistent with zero within 
the precision of numerical calculation.

\RconstSixthTable

We should like to emphasize that
the development of automatic code generator 
\cite{Aoyama:2005kf,Aoyama:2007IR}
was crucial in discovering
the existence of extra IR subtraction terms
in $M_{16}$ and $M_{18}$.
Details of our investigation will be reported elsewhere 
\cite{Aoyama:2007M8}. 
Adding the terms Eq.~(\ref{M16_add}) and Eq.~(\ref{M18_add}) to
the ``old" calculation Eq.~(58) of Ref.~\cite{Kinoshita:2005zr}, we find the entire contribution of Group V:
\begin{equation}
A_{1}^{(8)}({\rm Group V})= -2.179~16 ~(343),
\label{newa8V}
\end{equation}
which is in good agreement with the {\it still tentative} 
value obtained by the code generated by ``gencodeN":
\begin{equation}
A_{1}^{(8) {\rm genN}} ( {\rm Group V}) = -2.205 ~(54).
\end{equation}

The revised contribution (\ref{newa8V}) shifts the total 
eighth-order term $A_1^{(8)}$ to the one  given in Eq.~(\ref{newa8}).
As a consequence, the theoretical prediction of $a_e$ 
is moved by  $-5.421~775~(62)\times 10^{-12}$, yielding 
\begin{eqnarray}
a_e({\rm Rb})&=&1~159~652~182.78~(7.72)(0.11)(0.26)\times 10^{-12},
\nonumber \\
a_e({\rm Cs})&=&1~159~652~172.98~(9.33)(0.11)(0.26)\times 10^{-12},
\nonumber \\
\label{theory_ae}
\end{eqnarray}
where 7.72 and 9.33 come from the uncertainties of
the input values of the fine structure constant 
\begin{eqnarray}
&&\alpha^{-1}({\rm Rb06}) =~137.035~998~84~(91) ~~ [ {\rm 6.7 ppb}],
\label{alphaRb}
\\
&&\alpha^{-1}({\rm Cs06}) =~137.036~000~00~(110)~[ {\rm 8.0 ppb}],~  
\label{alphaCs}
\end{eqnarray}  
determined by the Rubidium atom \cite{Clade:2006} and
Cesium atom \cite{Wicht:2002,Gerginov:2006} experiments, respectively.
The uncertainty 0.11 of Eq.~(\ref{theory_ae}) comes from 
the eighth-order calculation
and 0.26 is an estimated uncertainty of the tenth-order term.

Because of high precision of the experiment (\ref{aeHV06value})
the fine structure constant $\alpha$ determined from the theory
and the measurement is sensitive to the revision of theory. 
The inverse fine structure constant $\alpha^{-1}(a_e)$ moves  
by $-6.411~80~(73) \times 10^{-7}$ from the previous value 
in Ref.~\cite{Gabrielse:2006gg}. The revised  $\alpha^{-1}(a_e07)$ is 
about 4.7 ppb (or about 7 s.~d.) smaller than 
$\alpha^{-1} (a_e06)$,  
but is still in good agreement with 
$\alpha^{-1}({\rm Rb06})$ of Eq.~(\ref{alphaRb})
and $\alpha^{-1}({\rm Cs06})$ of Eq.~(\ref{alphaCs}),
whose uncertainties are about 7 ppb.

\begin{acknowledgments} 
This work is supported in part by the JSPS
Grant-in-Aid for Scientific Research (C)19540322. 
T. K.'s work is supported by the U. S. National Science Foundation
under Grant No. PHY-0355005.
Numerical calculations were 
conducted on the RIKEN Super Combined Cluster System(RSCC).
\end{acknowledgments}


\end{document}